\documentclass[10pt,journal, final]{IEEEtran}
\usepackage{ifpdf}
\usepackage{cite}
\usepackage{algorithmic}
\usepackage{algorithm}

\ifCLASSINFOpdf
\usepackage[pdftex]{graphicx}

\else
 \usepackage[dvips]{graphicx}
\fi
\usepackage[cmex10]{amsmath}
\usepackage{epstopdf}
\usepackage{array}
\usepackage{flushend}
\usepackage{balance}
\usepackage{eqparbox}

\usepackage[tight,footnotesize]{subfigure}
\usepackage{fixltx2e}
\usepackage{color}
\usepackage{float}

\usepackage{dblfloatfix}
\usepackage{url}
\usepackage{cite}
\usepackage{subfigure}
\usepackage{amsmath}
\usepackage{epstopdf}

\newtheorem{lemma}{Lemma}

\usepackage{amssymb}
\usepackage{color}

\begin{document}
\allowdisplaybreaks[3]
\title{Uplink MIMO Performance Analysis for Diverse HAPS Antenna Array Architectures}
\setlength{\columnsep}{0.21 in}

\author{Shasha~Liu,~\IEEEmembership{Student Member,~IEEE,}
Abla~Kammoun,~\IEEEmembership{Member,~IEEE,}      
and
Mohamed-Slim~Alouini,~\IEEEmembership{Fellow,~IEEE}


}
\maketitle
\begin{abstract}
High-altitude platform stations (HAPS) are promising components of 6G and beyond networks, where antenna array configuration is critical for achieving wide-area coverage and high capacity with massive MIMO. This paper investigates and compares the uplink signal-to-interference-plus-noise ratio (SINR) distributions of user equipments (UEs) for five antenna array structures, including the cylindrical antenna array, the 3GPP antenna array, the hemispherical antenna array, and two proposed architectures, namely the truncated cone and truncated hemispherical antenna arrays, under uniform, Gaussian, and Poisson cluster process UEs distributions. Simulation results show that both proposed arrays achieve performance comparable to the hemispherical array, with the truncated hemispherical array being particularly effective for densely distributed UEs, while the truncated cone array offers a favorable tradeoff between performance and implementation complexity.
\end{abstract}
\IEEEpeerreviewmaketitle
\section{Introduction}
\subsection{Background}
High-altitude platform stations (HAPS) have attracted significant attention as key components of sixth-generation (6G) and beyond in wireless networks \cite{alam2021high}. Operating as aerial base stations, HAPS can effectively complement terrestrial networks by extending seamless coverage to regions where ground-based infrastructure is impractical, such as oceanic and mountainous areas. Typically deployed in the stratosphere at an altitude of approximately $20$ km, HAPS are capable of maintaining a quasi-stationary position relative to the Earth. Such deployment enables stable line-of-sight (LoS) communication links and wide-area coverage, while offering lower path loss and reduced latency compared with satellite-based systems.

HAPS can enhance service-link capacity by serving a large number of user equipments (UEs) simultaneously within its coverage area. To this end, massive multiple-input multiple-output (mMIMO) has been considered for HAPS systems, where large-scale antenna arrays enable multiuser MIMO (MU-MIMO) transmission with narrow, user-specific beamforming. This spatial multiplexing capability significantly increases the service-link capacity.

Antenna configuration plays a crucial role in determining both coverage area and capacity for HAPS \cite{tashiro2021cylindrical,abbasi2024hemispherical,maki2025downtilt}.
For example, cylindrical arrays can provide wide-area coverage of up to approximately 100 km, whereas conventional planar array–based systems are typically limited to coverage radius of 20–60 km \cite{tashiro2021cylindrical}.
Therefore, selecting an appropriate antenna configuration that matches the target coverage size and UE distribution is essential.
Motivated by this observation, this paper investigates and compares different antenna array configurations under varying coverage sizes and UE distribution scenarios.

\subsection{Related Work}
In HAPS literature, two main antenna types have been proposed for ground cell formation: aperture-type antennas \cite{thornton2003optimizing} and array-type antennas \cite{el2002cellular}. Aperture-type antennas generate a single beam per antenna, requiring multiple mechanically steered antennas to cover different cells. In contrast, array-type antennas enable electronic beamforming, allowing multiple beams to be simultaneously steered toward different ground locations. With a large number of antenna elements, high array gain can be achieved. 
\par
The antenna array for HAPS can be configured using various architectures. Most early studies on mMIMO systems for HAPS assume a linear or planar antenna array mounted on the underside of the platform, with all antenna elements oriented toward the Earth’s surface. Such a configuration primarily benefits UEs located directly beneath the HAPS. However, it provides negligible antenna gain for UEs at large horizontal distances. To overcome this limitation, \cite{tashiro2021cylindrical} proposed a cylindrical antenna array (CAA), in which antenna elements are deployed on both the cylindrical side surface and the bottom panel of the platform to enhance coverage for distant UEs. This architecture has been shown to improve both coverage and capacity. Nevertheless, the proposed structure overlooks the importance of antenna tilt angles,  and the elements on the cylindrical surface have boresights parallel to the ground, resulting in a loss of antenna gain in the zenith direction.
Further, the 3GPP study in \cite{3gpp_ntn_rf_r18} introduced a hexagonal antenna array comprising a downward-facing panel serving the center cell and six outward-facing panels serving the surrounding cells. Such a structure considers the tilt angle for the side panels. However, it only has six panels, which limits the freedom of azimuth angle.
To overcome these limitations, we propose a truncated cone antenna array, in which the elements deployed on the curved surface are tilted to avoid the loss of antenna gain in the zenith direction while preserving the freedom of azimuth angle, thereby enhancing antenna gain for UEs. In addition, antenna elements are deployed on the bottom panel to guarantee the performance of UEs directly beneath the HAPS.
The authors in \cite{abbasi2024hemispherical} also recognized the importance of ensuring that each UE is directly aligned with specific antenna elements. Accordingly, the hemispherical antenna array was proposed in \cite{abbasi2024hemispherical}. However, this structure provides limited antenna gain for UEs located directly beneath the HAPS. To address this issue, we further propose a truncated hemispherical antenna array that incorporates an additional circular panel at the bottom to enhance coverage for nadir UEs.
\par
This paper compares the signal-to-interference-plus-noise ratio (SINR) distributions of UEs in the uplink for five antenna array structures under uniform, Gaussian, and archipelago-like Poisson cluster process UE distributions. Simulation results show that the proposed truncated cone and truncated hemispherical arrays achieve performance comparable to the HAA, with the truncated hemispherical array being particularly suitable for densely distributed UEs.

\section{Configuration of Antenna Arrays}
\begin{figure*}[t]
  \centering
  \subfigure[Cylindrical array\label{Fig: AN_cylindrical}]{
    \includegraphics[width=0.3\linewidth]{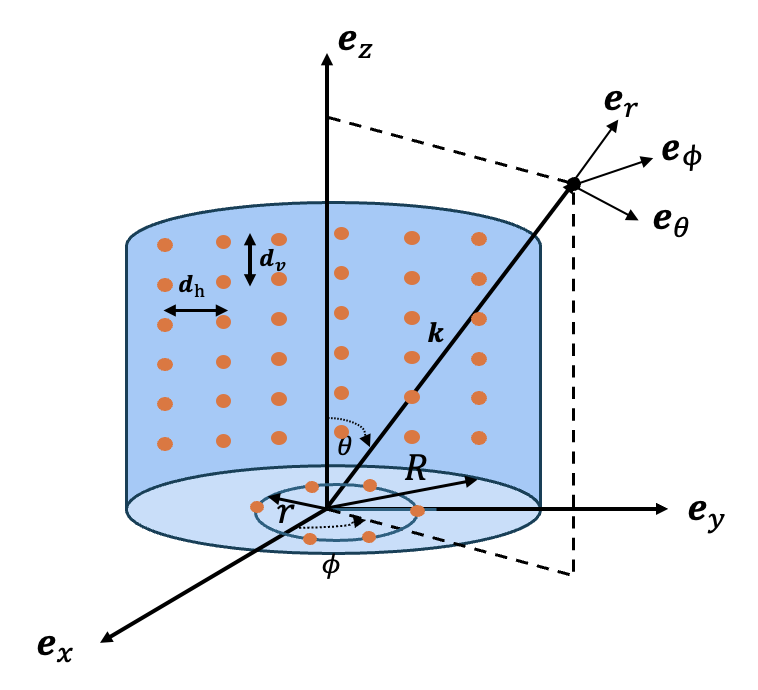}
  }\hfill
  \subfigure[3GPP array\label{Fig: AN_3GPP}]{
    \includegraphics[width=0.35\linewidth]{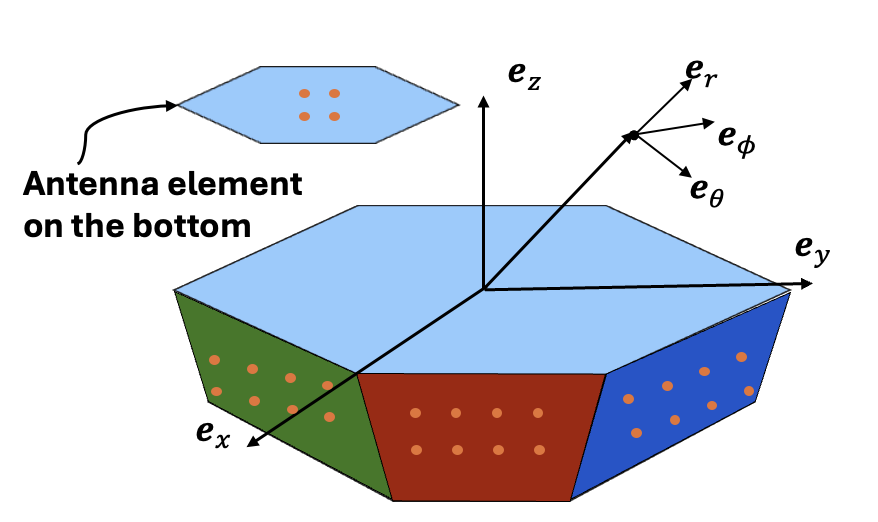}
  }\hfill
  \subfigure[Hemispherical array\label{Fig: AN_Hemispherical}]{
    \includegraphics[width=0.3\linewidth]{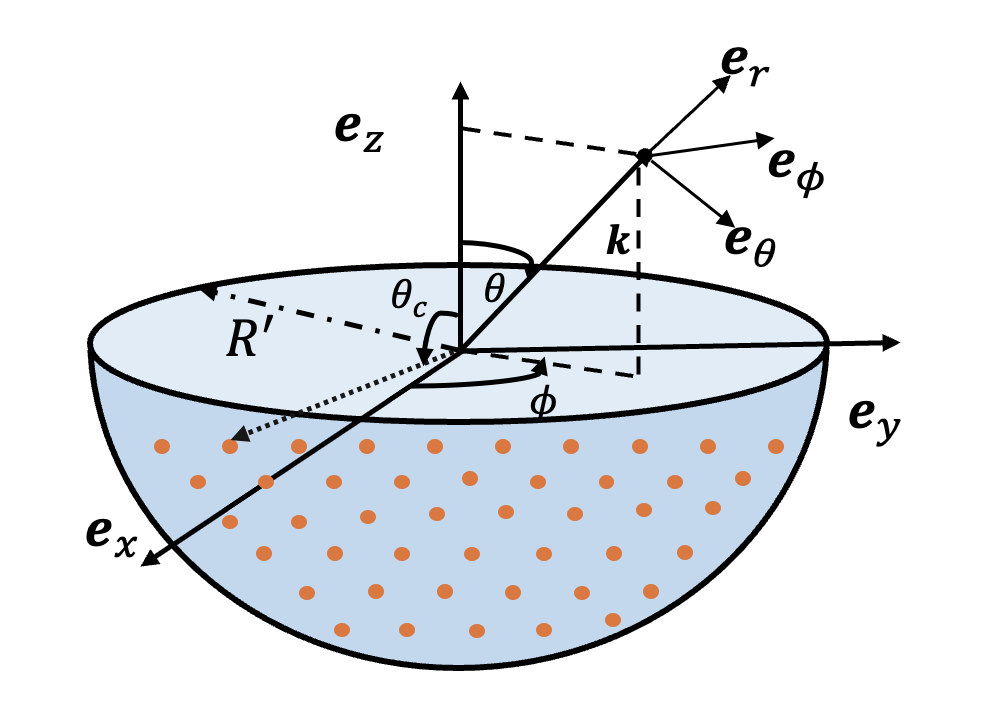}
  }

  \vspace{2mm} 

  \subfigure[Truncated cone array\label{Fig: AN_Tx_cone}]{
    \includegraphics[width=0.3\linewidth]{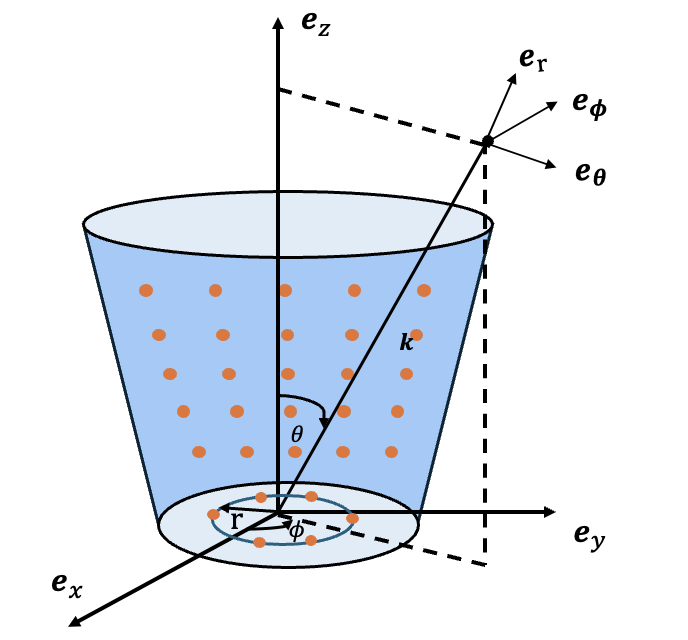}
  }\hspace{4mm}
  \subfigure[Truncated hemispherical array\label{Fig: AN_truncated_Hemispherical}]{
    \includegraphics[width=0.3\linewidth]{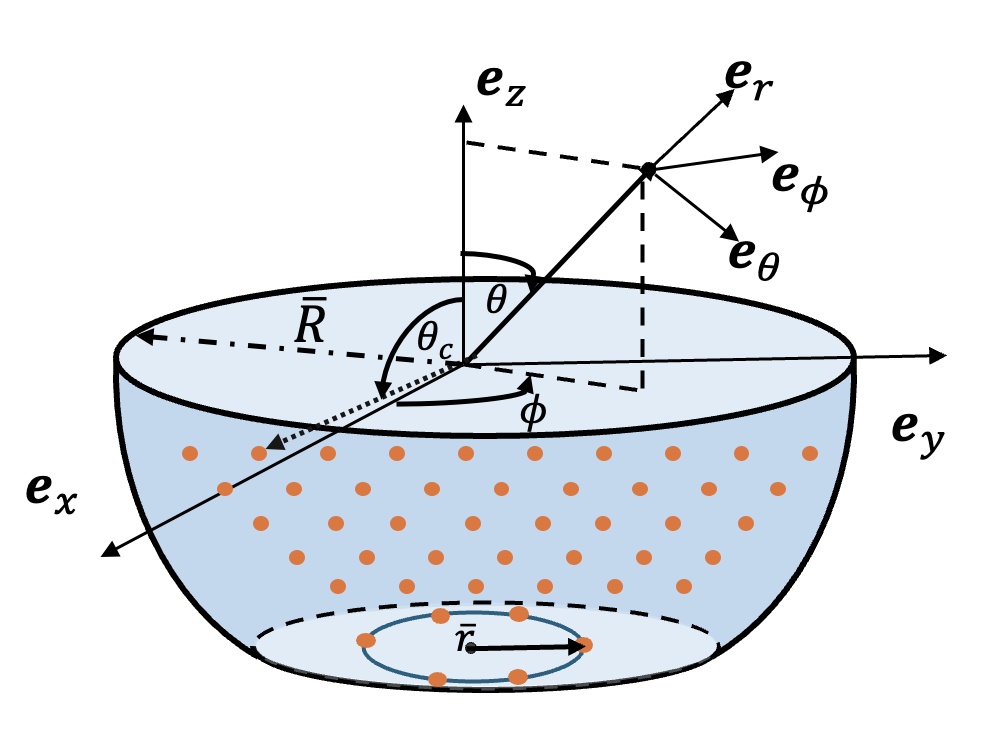}
  }

  \caption{Configuration of antenna arrays}
  \label{Fig:antenna_arrays}
\end{figure*}

In this section, as illustrated in Fig.~\ref{Fig:antenna_arrays}, the five different antenna arrays are introduced. Detailed descriptions of these antenna arrays are presented in the following subsections.
\subsection{Cylindrical Array Antenna}
Fig.~\ref{Fig: AN_cylindrical} illustrates the cylindrical array.
On the curved surface, $N_h$ antenna elements are uniformly distributed along
the azimuth with spacing $d_h$ on a circle of radius $R$, while $N_v$ elements
are vertically arranged with spacing $d_v$.
To provide coverage for the region directly below the HAPS, an additional
$N_b$ antenna elements are uniformly deployed on the bottom face of the array.

\subsubsection{Antenna Element on the curved surface}
\label{subsucsection: bottom_sylindrical}
Let $s=1,\ldots,N_h$ and $z=1,\ldots,N_v$ denote the azimuthal column index and
vertical ring index, respectively, with
$\hat{\psi}_s=\frac{2\pi(s-1)}{N_h}$.
Then, the horizontal radiation pattern of the element located in the $s$-th column and 
$z$-th ring is expressed as
$
g_{E,H}^{(z,s)}(\phi) = g_{E,H}(\phi - \hat{\psi}_s),
$
Accordingly, its  overall radiation pattern with respect to the global coordinate system is given by:
\begin{equation}
F_{\rm{cyl}}^{(z,s)}(\phi,\theta):= g_{E}^{(z,s)}(\phi, \theta)=g_{E,H}(\phi-\hat{\psi}_s)g_{E,V}(\theta)
\end{equation}
The array response of the $(z,s)$-th element is expressed as
\begin{equation}
\label{Eq:V_C}
    [\mathbf{V}]_{z,s}^{\rm cyl}=\exp[ik(R\cos(\phi-\hat{\psi}_s)\sin \theta \nonumber+(z-1)d_v\cos \theta)],
\end{equation}
which follows from the element position
$\mathbf x_{z,s}=(R\cos\hat{\psi}_s,\,R\sin\hat{\psi}_s,\,(z-1)d_v)$
and the incident direction
$\hat{\mathbf v}=(\cos\phi\sin\theta,\,\sin\phi\sin\theta,\,\cos\theta)$.
\subsubsection{Antenna elements mounted on the bottom face of the cylinder}
The bottom elements form a uniform circular array of radius $r$ and are oriented
toward the direction $-{\bf e}_z$, which can be equivalently modeled by rotating
each element about the ${\bf e}_y$ axis by $\beta=\pi/2$. 
Therefore, let $F^{(b)}(\phi,\theta)$ denote the radiation pattern of an antenna element mounted on the bottom face, evaluated in the direction specified by the azimuth angle $\phi$ and elevation angle $\theta$ with respect to the global coordinate system.
Then,
\begin{equation}
    F_{\rm cyl}^{(b)}(\phi,\theta) = g_E(\phi_a, \theta_a),
\end{equation}
where $\theta_a= \operatorname{arcos}(\cos \phi \sin \theta)$ and $\phi_a =\tan^{-1} \left(\frac{\sin\phi\sin\theta}{-\cos \theta}\right)$.
For elements in the bottom,  we have the array response  
\begin{equation}
\label{Eq:V_B}
[\tilde{\mathbf{v}}(r,N_b)]_{b}^{\rm cyl}= \exp{\left(ikr\cos(\phi-\psi_b)\sin \theta \right)},
\end{equation}
where $\psi_b=\frac{2\pi(b-1)}{N_b}, b=1,\cdots, N_b$. 
\subsubsection{Overall array response and pattern}
Let $\mathcal{S}_{\mathrm{cyl}}$ and $\mathcal{S}_{\mathrm{bot}}$ denote the index sets of antenna elements located on the curved surface and on the bottom face of the cylindrical array, respectively.
For a plane wave arriving from direction $(\phi, \theta)$, the array response and
the radiation pattern are
\begin{equation}
[{\bf a}_{\mathrm{tx}}^{\text{cyl}}(\phi,\theta)]_q=
\begin{cases}
{\bf V}_{z,s}^{\rm cyl},      & \text{if } q\in \mathcal{S}_{\mathrm{cyl}},\\[2pt]
[\tilde{\bf v}]_b^{\rm cyl},  & \text{if } q\in \mathcal{S}_{\mathrm{bot}}.
\end{cases}
\label{eq:array_response}
\end{equation}
\begin{equation}
F_{\rm cyl}^q(\phi,\theta)=
\begin{cases}
F_{\rm cyl}^{(z,s)}(\phi,\theta), & \text{if } q\in \mathcal{S}_{\mathrm{cyl}},\\[2pt]
F_{\rm cyl}^{(b)}(\phi,\theta),   & \text{if } q\in \mathcal{S}_{\mathrm{bot}},
\end{cases}
\label{eq:pattern}
\end{equation}
\subsection{3GPP antenna structure}
We consider the hexagonal antenna array specified in \cite{3gpp_ntn_rf_r18}.
The hexagonal antenna array comprises seven antenna panels, including six side panels and one downward-facing panel, as illustrated in Fig.~\ref{Fig: AN_3GPP}. The corresponding antenna parameters are specified in Table~6.2.3.3-1 of \cite{3gpp_ntn_rf_r18}. The bottom panel is tilted by $90^{\circ}$ with respect to the horizontal plane, while each side panel is tilted by $\theta_{\rm tilt}=23^\circ$.
\subsubsection{Antenna elements on side-panel}
Each side panel consists of a $P_{\text{curve}} \times Q_{\text{curve}}$ antenna elements, with an inter-element spacing of $d_p$.  
Panel $p$, is first rotated about the $z$-axis by an angle 
$\alpha = \phi_p= \frac{(p-1)\pi}{3}$, followed by a rotation about the $y$-axis by an angle $\beta = \theta_{\text{tilt}}$.
Following lemma \ref{lemma: GCS2LCS}, the radiation pattern of the antenna element
located at the $m$-th row and $n$-th column on panel $p$ is
\begin{equation}
    F^{(p,m,n)}_{\rm 3gpp}(\phi,\theta) = g_E(\phi_a, \theta_a),
\end{equation} 
where $\theta_a=\arccos \left(\sin \theta \sin \theta_{\text {tilt }} \cos \left(\phi-\phi_p\right)+\cos \theta \cos \theta_{\text {tilt }} .\right)$ and $\phi_a = \tan^{-1}\left(\frac{\sin \theta \sin (\phi-\phi_p)}{\sin\theta \cos \theta_{\text{tilt}} \cos(\phi-\phi_p)-\cos \theta \sin \theta_{\text{tilt}}}\right)$.
The position of the $(m,n)$-th element on panel $p$ in the GCS is given
\begin{equation}
   \hat{\mathbf{x}}_{m,n}^p = \mathbf{x}_{m,n}\mathbf{R}(p)^{-1},
\end{equation}
where 
$
\mathbf{x}^p_{m,n}=\left( \frac{\sqrt{3}l}{2},\; -\frac{(P_{\text{curve}}-1)d_h}{2} + (n-1)d_h,\; -(m-1)d_h \right), 
$ with 
$l = (P_{\text{curve}}+1)d_h$ and
$\mathbf{R}(p)$ denotes the rotation matrix in lemma \ref{lemma: GCS2LCS}.
The array response element corresponding to the antenna element $(m,n)$ on the panel $p$ can be written as
\begin{equation}
[\mathbf{V}]_{m,n}^{\rm 3gpp}=\exp(ik\hat{\mathbf{v}}\cdot\hat{\mathbf{x}}_{m,n}^p)
\end{equation}
\subsubsection{Antenna elements on the bottom-panel}
The bottom panel consists of a
$P_{\rm bottom}\times Q_{\rm bottom}$ planar array whose boresight is oriented
toward $-{\bf e}_z$.
The radiation pattern of the $(m,n)$-the bottom element is modeled as
\begin{equation}
    F^{(b,m,n)}_{\rm 3gpp}(\phi,\theta) = g_E(\phi_a, \theta_a),
\end{equation}
where $(\phi_a, \theta_a)$ are defined in the corresponding LCS, which follows \ref{subsucsection: bottom_sylindrical}.  
\par
The array response element corresponding to the $(m,n)$-th antenna element can be written as
\begin{equation}
[\tilde{\mathbf{V}}]_{m,n}^{\rm 3gpp}=\exp(ik\hat{\mathbf{v}}\cdot{\mathbf{x}}_{m,n})
\end{equation}
with the position of the $(m,n)$-th bottom element is
$
\mathbf{x}_{m,n}=( -\frac{(Q_{\text{bottom}}-1)d_h}{2} + (m-1)d_h,\; -\frac{(P_{\text{bottom}}-1)d_h}{2} + (n-1)d_h,\ d_{\text{min}} ),
$
where $d_{\rm min}$ denotes the minimum $z$-coordinate among the side-panel.
\subsubsection{Overall array response and radiation pattern}
Let $\mathcal{S}_{\mathrm{side}}$ and $\hat{\mathcal{S}}_{\mathrm{bot}}$ denote the index sets of antenna elements
located on the six side panels and on the bottom panel of the 3GPP hexagonal array, respectively. we have
\begin{equation}
[{\bf a}_{\rm tx}^{\rm 3gpp}(\phi,\theta)]_q=
\begin{cases}
[\mathbf{V}]_{m,n}^{\rm 3gpp}(p;\phi,\theta), & q\in\mathcal{S}_{\rm side},\\[3pt]
[\tilde{\mathbf{V}}]_{m,n}^{\rm 3gpp}(\phi,\theta), & q\in\hat{\mathcal{S}}_{\rm bot}.
\end{cases}
\label{eq:3gpp_array_response}
\end{equation}
\begin{equation}
F_q^{\rm 3gpp}(\phi,\theta)=
\begin{cases}
F_{\rm 3gpp}^{(p,m,n)}(\phi,\theta), & q\in\mathcal{S}_{\rm side},\\[3pt]
F_{\rm 3gpp}^{(b,m,n)}(\phi,\theta), & q\in \hat{\mathcal{S}}_{\rm bot}.
\end{cases}
\label{eq:3gpp_radiation_pattern}
\end{equation}

\subsection{Hemispherical Antenna Array}
\label{subsection: Hemisperical}
Considering the HAA proposed in \cite{abbasi2024hemispherical}, as shown in Fig.~\ref{Fig: AN_Hemispherical}.
The array consists of $N$ vertically stacked circular rings.
The $c$-th ring has radius $r_c$ and contains $N_c$ uniformly spaced antenna
elements with inter-element spacing $d_u$.
The inter-ring spacing is $d_c$, and the hemispherical radius is $R^{\prime}$.
The $u$-th antenna element on the $c$-th ring is described by the spherical
coordinates $(R',\theta_c,\phi_u)$, where
$\phi_u=\frac{2\pi(u-1)}{N_c}$.
The radiation pattern of element $(c,u)$ is expressed as
\begin{equation}
    F^{(c,u)}_{\rm hemi}(\phi,\theta) = g_E(\phi_a, \theta_a),
\end{equation} 
where $\theta_a=\arccos \left(-\cos \theta_c \sin \theta \cos \left(\phi-\phi_u\right)+\sin \theta_c \cos \theta\right)$ and $\phi_a =\tan^{-1} \left(\frac{\sin \theta \sin \left(\phi-\phi_u\right)}{\sin \theta \sin \theta_c \cos \left(\phi-\phi_u\right)+\cos \theta \cos \theta_c}\right)$.
This coordinate transformation corresponds to a rotation by $\alpha=\phi_u$ about the
$z$-axis, followed by a rotation by $\beta=\theta_c-\frac{\pi}{2}$ about the $y$-axis.
\par
Let the departure direction be
$\hat{\mathbf v}=(\sin\theta\cos\phi,\ \sin\theta\sin\phi,\ \cos\theta)$,
and the position of element $(c,u)$ be
$
\mathbf x_{c,u}=
(r_c\cos\phi_u,\ r_c\sin\phi_u,\ -(c-1)d_c).
$
The corresponding array response is given by
\begin{equation}
[\mathbf{V}(R)]_{c,u}^{\rm hemi}
=\exp\!\left(jk\!\left[
r_c\sin\theta\cos(\phi-\phi_u)-(c-1)d_c\cos\theta
\right]\right).
\end{equation}
Let $\mathcal{S}_{\mathrm{hemi}}$ denote the index set of all transmit antenna elements. We get
$
[{\bf a}_{\mathrm{tx}}^{\mathrm{hemi}}(\phi,\theta)]_{q}
= [\mathbf{V}]_{c,u}^{\rm hemi}, \  q \in \mathcal{S}_{\mathrm{hemi}}
$ and 
$
F_q^{\rm hemi}(\phi,\theta)=F^{(c,u)}_{\rm hemi}, \ q \in \mathcal{S}_{\mathrm{hemi}}
$

\subsection{Truncated Cone Array }
To investigate the impact of the element tilt angle on the cylindrical array,
we introduce a truncated cone antenna structure by tilting the elements on the
curved surface by an angle $\hat{\theta}_{\rm tilt}$ with respect to the horizontal
plane, as illustrated in Fig.~\ref{Fig: AN_Tx_cone}.
\subsubsection{Antenna elements on the truncated cone surface}
The antenna element indexed by $(s,z)$ is first rotated about the $z$-axis by
$\hat{\psi}_s=\frac{2\pi(s-1)}{N_h}$ and then tilted about the $y$-axis by
$\hat{\theta}_{\rm tilt}$.
Accordingly, the radiation pattern of the $(s,z)$-th element is given by
\begin{equation}
    F^{(s,z)}(\phi,\theta) = g_E(\phi_a, \theta_a),
\end{equation}
where $\theta_a
= \arccos \left(\sin \theta \sin \hat{\theta}_{\text {tilt }} \cos \left(\phi-\phi_s\right)+\cos \theta \cos \hat{\theta}_{\text {tilt }} .\right) $ and $\phi_a =\tan^{-1}\left(\frac{\sin \theta \sin (\phi-\phi_s)}{\sin\theta \cos \hat{\theta}_{\text{tilt}} \cos(\phi-\phi_s)-\cos \theta \sin \hat{\theta}_{\text{tilt}}}\right)$.
\par
Hence, the array response element corresponding to the $(s,z)$-th antenna element can be written as
\begin{equation}
[\mathbf{V}]_{s,z}^{\rm cone}=\exp(ik\hat{\mathbf{v}}\cdot\hat{\mathbf{x}}_{s,z})
\end{equation}
with the location of the antenna element is
$
\hat{\mathbf{x}}_{s,z} = \mathbf{x}_{z,s}\hat{\mathbf{R}}^{-1},
$
where $\mathbf{x}_{s,z}=(R, 0, (z-1)d_v)$ and $\hat{\mathbf{R}}$ denotes the rotation matrix in lemma \ref{lemma: GCS2LCS}. 
\subsubsection{Antenna elements on the bottom face}
The antenna elements on the bottom face follow the same configuration as in
Section~\ref{subsucsection: bottom_sylindrical}.
Thus, the radiation pattern and array response satisfy
$F_{\rm cone}^{(b)}(\phi,\theta)=F_{\rm cyl}^{(b)}(\phi,\theta)$ and
$[\tilde{\mathbf v}]_{b}^{\rm cone}=[\tilde{\mathbf v}]_{b}^{\rm cyl}$.
\subsubsection{Overall array response and radiation pattern}
Let $\mathcal{S}_{\mathrm{cone}}$ and $\mathcal{S}_{\mathrm{bot}}$ denote the index sets of antenna elements located on the curved surface and on the bottom face of the truncated cone array, respectively.
The array response and the radiation pattern are
\begin{equation}
[{\bf a}_{\mathrm{tx}}^{\text{cone}}(\phi,\theta)]_q=
\begin{cases}
{\bf V}_{z,s}^{\rm cone},      & \text{if } q\in \mathcal{S}_{\mathrm{cone}},\\[2pt]
[\tilde{\bf v}]_b^{\rm cone},  & \text{if } q\in \mathcal{S}_{\mathrm{bot}}.
\end{cases}
\label{eq:array_response}
\end{equation}
\begin{equation}
F_{\rm cone}^q(\phi,\theta)=
\begin{cases}
F_{\rm cone}^{(z,s)}(\phi,\theta), & \text{if } q\in \mathcal{S}_{\mathrm{cone}},\\[2pt]
F_{\rm cone}^{(b)}(\phi,\theta),   & \text{if } q\in \mathcal{S}_{\mathrm{bot}},
\end{cases}
\label{eq:pattern}
\end{equation}

\subsection{Truncated hemispherical Array Antenna}
To enhance the coverage and performance for UEs located directly beneath
the HAPS, we propose a truncated hemispherical antenna array, which
extends the conventional hemispherical array by incorporating an additional
bottom circular panel, as illustrated in
Fig.~\ref{Fig: AN_truncated_Hemispherical}.
Similar to the hemispherical array described in
Section~\ref{subsection: Hemisperical}, the THS consists of $\bar{N}$ parallel
circular rings stacked along the vertical axis.
The $c$-th ring contains $\bar{N}_c$ uniformly spaced antenna elements along
its circumference, with a hemispherical radius $\bar{R}$.
In addition, $\bar{N}_b$ antenna elements are uniformly deployed on a bottom
circular panel of radius $\bar{r}$.
\par
\subsubsection{Antenna elements on the truncated hemispherical surface}
For the antenna elements located on the curved surface, the radiation pattern
and array response are identical to those of the conventional hemispherical
array, i.e.,
$
F_{\rm THS}^{(c,u)}(\phi,\theta)
=F_{\rm hemi}^{(c,u)}(\phi,\theta),
$
$
[\mathbf V]_{c,u}^{\rm THS}
=[\mathbf V(\bar{R})]_{c,u}^{\rm hemi}.
$
\subsubsection{Antenna elements on the bottom panel}
The bottom-panel elements form a uniform circular array of radius $\bar{r}$.
Following Section~\ref{subsucsection: bottom_sylindrical}, the radiation pattern
and array response of the $b$-th bottom element are given by
$
F_{\rm THS}^{(b)}(\phi,\theta)
=F_{\rm cyl}^{(b)}(\phi,\theta),
$
$
[\tilde{\mathbf v}]_{b}^{\rm THS}
=[\tilde{\mathbf v}(\bar{r},\bar{N}_b)]_{b}^{\rm cyl}.
$
\subsubsection{Overall array response and radiation pattern}
Let $\mathcal{S}_{\mathrm{THS}}$ and $\bar{\mathcal{S}}_{\mathrm{bot}}$ denote the index sets of antenna elements located on the curved surface and on the bottom face of the cylindrical array, respectively.
The array response and the radiation pattern are
\begin{equation}
[{\bf a}_{\mathrm{tx}}^{\rm THS}(\phi,\theta)]q =
\begin{cases}
{\bf V}_{c,u}^{\rm THS}, & \text{if } q \in \mathcal{S}_{\mathrm{THS}}, \\
[\tilde{\bf v}]_b^{\rm THS}, & \text{if } q \in \bar{\mathcal{S}}_{\mathrm{bot}} .
\end{cases}
\label{eq:array_response_THS}
\end{equation}

\begin{equation}
F_{\rm THS}^q(\phi,\theta) =
\begin{cases}
F_{\rm THS}^{(c,u)}(\phi,\theta), & \text{if } q \in \mathcal{S}_{\mathrm{THS}}, \\
F_{\rm THS}^{(b)}(\phi,\theta), & \text{if } q \in \bar{\mathcal{S}}_{\mathrm{bot}} .
\end{cases}
\label{eq:pattern_THS}
\end{equation}

\section{Channel Model}
Consider a wireless channel composed of \( N_{cl} \) scattering clusters. For the \( n \)-th cluster, let \( \phi_n \) and \( \theta_n \) denote angles of departure (AoDs) and the zenith angles
of departure (ZoDs), respectively. The channel contribution from the \( n \)-th cluster is modeled as:
\begin{equation}
\mathbf{h}_n=\sqrt{P_n}\sqrt{g_E\left(\phi_n, \theta_n \right)}\exp(j\Phi_n^{\theta\theta})\mathbf{a}_{tx}^H(\phi_n, \theta_n)
\end{equation}
where  \( P_n \) denotes the power associated with the \( n \)-th cluster, \( g_E(\phi_n, \theta_n) \) is the directional antenna gains at the transmitter, $\Phi_n^{\theta\theta}$ is the initial phase, which are uniformly distributed in the range $(-\pi,\pi)$, and  \( \mathbf{a}_{\mathrm{tx}}(\phi_n, \theta_n) \) represents the array response vectors of the transmitter.
In particular, the $q$-th entry of $\mathbf a_{\rm tx}(\phi,\theta)$ and the
associated radiation pattern are given by the corresponding
$\big[{\bf a}_{\rm tx}(\phi,\theta)\big]_q$ and $F_q(\phi,\theta)$ defined for
each array configuration in the previous sections.
\par
 The overall channel matrix \( \mathbf{h} \) follows a Rician fading model, given by
\begin{equation}
\mathbf{h}=\sqrt{\beta_l}(\sqrt{\frac{1}{K_R+1}}\mathbf{h}_{\text{NLoS}} + \sqrt{\frac{K_R}{K_R+1}}\mathbf{h}_{\text{LoS}})
\end{equation}
where 
$
\mathbf{h}_{\text{NLoS}}=\sum_{n=1}^{N_{cl}} \mathbf{h}_n
$
and
$$
\mathbf{h}_{\text{LoS}}=\exp(-j2\pi\frac{d_{3D}}{\lambda})\sqrt{g_E\left(\phi^{LoS}, \theta^{LoS} \right)}\mathbf{a}_{tx}^H
$$
and $\beta_l$ is calculated in dB scale as:
\begin{equation}
    \beta_{l,\mathrm{dB}} =P_t + G_{\mathrm{UE}} - PL_{l,\mathrm{dB}} + G_{E,\mathrm{max}},
\end{equation}
where $P_t$ denotes the UE's transmit power, $G_{\mathrm{UE}}$ is the UE's antenna gain, $PL_{l,\mathrm{dB}}$ represents the free-space path loss in dB, and $G_{E,\mathrm{max}}$ is the maximum element gain of the transmit antenna array.
Accordingly, the large-scale fading factor in linear scale is given by $\beta_l = 10^{\beta_{l,\mathrm{dB}} / 10}$.

\section{Performance Optimization}
\label{section: optimization}
Consider a uplink multi-user MISO system, where the HAPS serves 
$K$ non-cooperating single-element users. Denote the decoding vectors and the data symbol for user $\ell$ as $\mathbf{p}_{\ell} \in \mathbb{C}^{N_t \times 1}$ and $s_{\ell} \in \mathcal{CN}(0,1)$ respectively.  The received signal for the HAPS is formulated as:
\begin{equation}
\mathbf{x}=\sum_{\ell=1}^{K}\mathbf{h}_\ell s_\ell+\mathbf{n}=\mathbf{H}\mathbf{s}+\mathbf{n},
\end{equation}
where $\mathbf{H}=[\mathbf{h}_1, \cdots, \mathbf{h}_K]$ and $\mathbf{s}=[s_1, \cdots, s_K]^{T}$, and $\mathbf{n} \in \mathcal{CN}(0,\sigma^2\mathbf{I}_{N_t})$ denotes additive white Gaussian noise with zero mean and variance $\sigma^2$.
The received signal for $\ell^{th}$ users is decoded by $\mathbf{p}_{l}$ can be written as:
\begin{equation}
\tilde{s}_{\ell}=\mathbf{p}_{\ell}^H\mathbf{h}_{\ell} s_{\ell}+\sum_{i\neq l}\mathbf{p}_{\ell}^H\mathbf{h}_i s_i + \mathbf{p}_{\ell}^H\mathbf{n}
\end{equation}
where $\mathbf{p}_{l}$ is designed as the MMSE receiver, which is written as
$
\mathbf{p}_{\ell}=(\mathbf{H}\mathbf{H}^H+\sigma^2\mathbf{I}_{N_t})^{-1}\mathbf{h}_{\ell}.
$
Therefore, the SINR is formulated as:
\begin{equation}
\label{Eq: SINR}
\rm{SINR}_{\ell}=\mathbf{h}_{\ell}^H(\mathbf{H}_{\ell}\mathbf{H}_{\ell}^H+\sigma^2\mathbf{I})^{-1}\mathbf{h}_{\ell}
\end{equation}
To characterize the SINR performance across users, we evaluate the empirical
cumulative distribution function (CDF) of the SINR, defined as
$$
\mathbb{E}\left[\frac{\#\{\rm{SINR}\leq \gamma_{\rm th} \}}{K}\right]
$$
where $\gamma_{\rm th}$ denotes the SINR threshold and
$\#\{\cdot\}$ denotes the cardinality of a set.

\section{Numerical Results}
\label{section: simulation}
We consider a single HAPS deployed at an altitude of $H = 20~\mathrm{km}$. The antenna configuration parameters are summarized in Table~\ref{table:antenna}. 
For a UE indexed by \(\ell\) with Cartesian coordinates
\((x_\ell, y_\ell, z_\ell)\), the corresponding zenith and azimuth angles are given by
$
(\theta_\ell, \phi_\ell)
=\left(\frac{\pi}{2}+\arctan\!\left(
\frac{H - z_\ell}{\sqrt{x_\ell^2 + y_\ell^2}}
\right),\;\arctan\!\left(\frac{x_\ell}{y_\ell}\right)\right).
$
The thermal noise power (in dBm) is calculated as
$
\sigma_{\mathrm{dBm}}^2= -174 + 10 \log_{10}(\mathrm{BW})+\mathrm{NF},
$
where the system bandwidth is \(\mathrm{BW} = 360~\mathrm{kHz}\) and the noise figure is
\(\mathrm{NF} = 9~\mathrm{dB}\). The UE transmit power is set to
\(P_t = 23~\mathrm{dBm}\), and the UE antenna gain is
\(G_{\mathrm{UE}} = 0~\mathrm{dBi}\).
Furthermore, the parameters associated with the angles of departure (AoDs) and the zenith
angles of departure (ZoDs) in the 3GPP channel model are adopted from~\cite{3gpp_tr38811}.
\begin{table}[!h]
\centering
\caption{Antenna Array Parameters}
\label{table:antenna}
\renewcommand{\arraystretch}{1.1}
\begin{tabular}{|c|c|}
\hline
\textbf{Parameter} & \textbf{Value} \\
\hline
$N_h,\;N_v,\;N_b$ & $32,\;6,\;8$ \\
\hline
$P_{\rm curve}\!\times\!Q_{\rm curve}$ & $4 \times 8$ \\
\hline
$P_{\rm bottom}\!\times\!Q_{\rm bottom}$ & $2 \times 4$ \\
\hline
$\bar{N}_b$ & $8$ \\
\hline
$d_h,d_v,d_b,d_p,d_c,d_u$ & $0.7\lambda$ \\
\hline
$R,\;R',\;\bar{R}$ & $53.48,\;65.37,\;63.65$ cm \\
\hline
$r,\;\bar{r}$ & $13.37$ cm \\
\hline
$\theta_{\rm tilt},\;\hat{\theta}_{\rm tilt}$ & $23^\circ$ \\
\hline
$\theta_{\rm 3dB},\;\phi_{\rm 3dB}$ & $65^\circ$ \\
\hline
$G_{E,\max}^{\rm Tx}$ & $8$ dBi \\
\hline
$f_c$ & $2$ GHz \\
\hline
\end{tabular}
\end{table}

\par
\begin{figure*}[!h]
\centering
\subfigure[$K=10,R_{\text{radius}}=100$ km]{
    \includegraphics[width=2.2in]{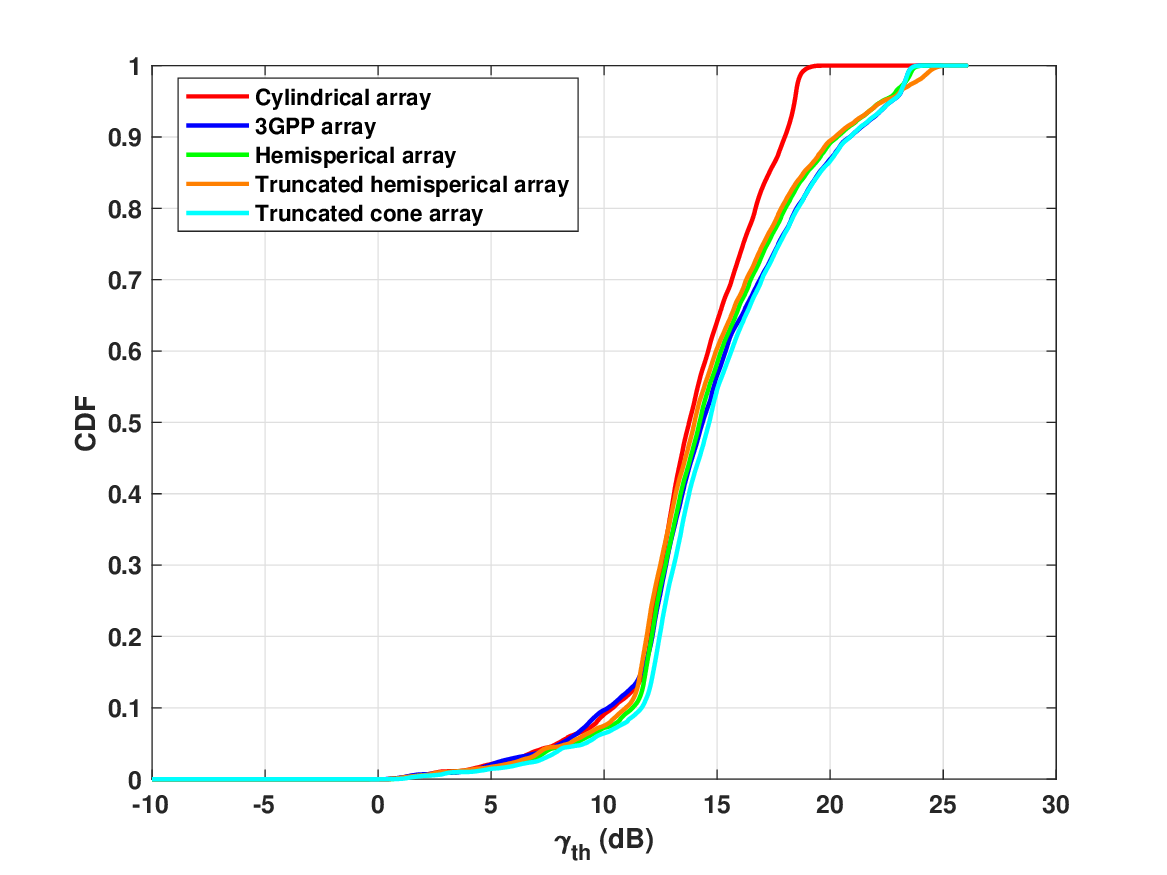}
    \label{Fig: RR100K10}
}
\subfigure[$K=20,R_{\text{radius}}=100$ km]{
    \includegraphics[width=2.2in]{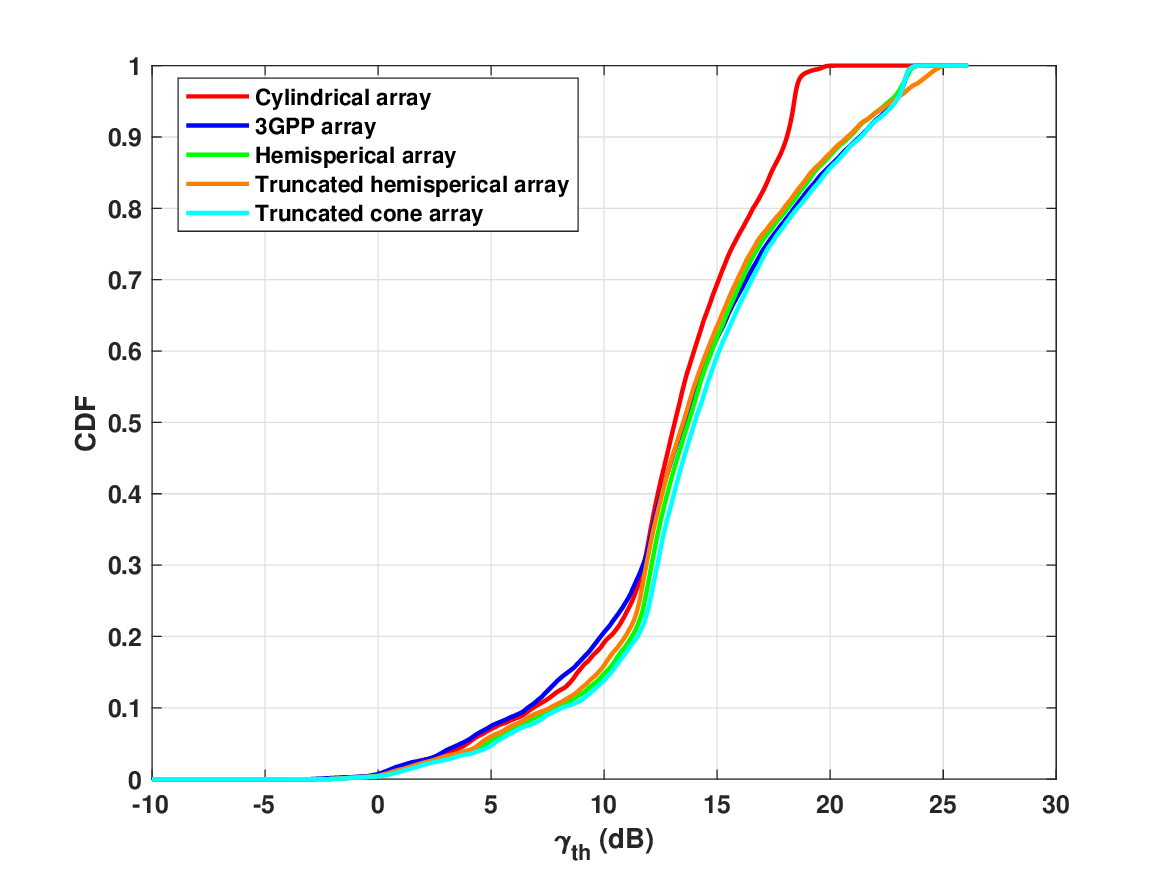}
    \label{Fig: RR100K20}
}
\subfigure[$K=50,R_{\text{radius}}=100$ km]{
    \includegraphics[width=2.2in]{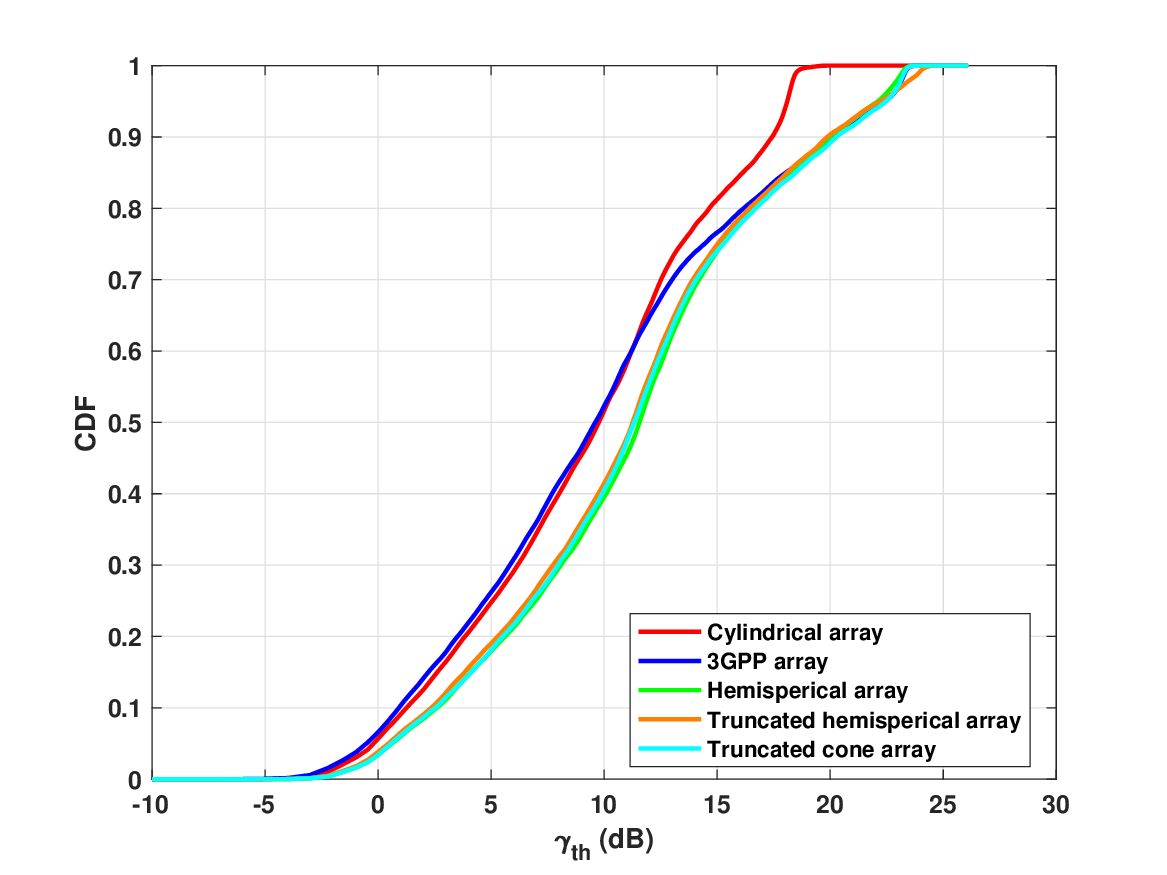}
    \label{Fig: RR100K50}
}
\caption{Uniformly distributed UEs in a circular cell of radius $R_{\text{radius}}=100$ km.}
\label{Fig:R100_uniform}
\end{figure*}
\par
\begin{figure*}[!h]
\centering
\subfigure[$K=10, R_{\text{radius}}=20$ km]{
    \includegraphics[width=2.2in]{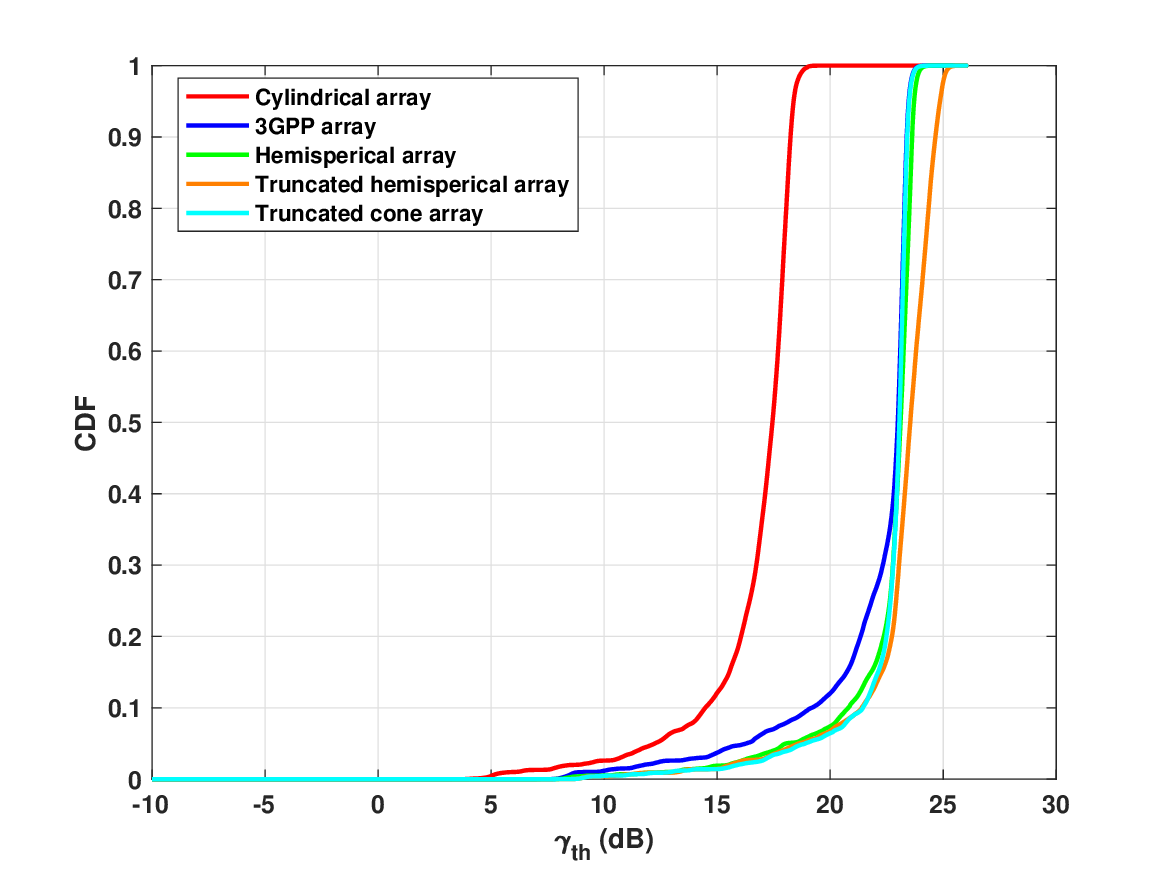}
    \label{Fig: RR20K10}
}
\subfigure[$K=20, R_{\text{radius}}=20$ km]{
    \includegraphics[width=2.2in]{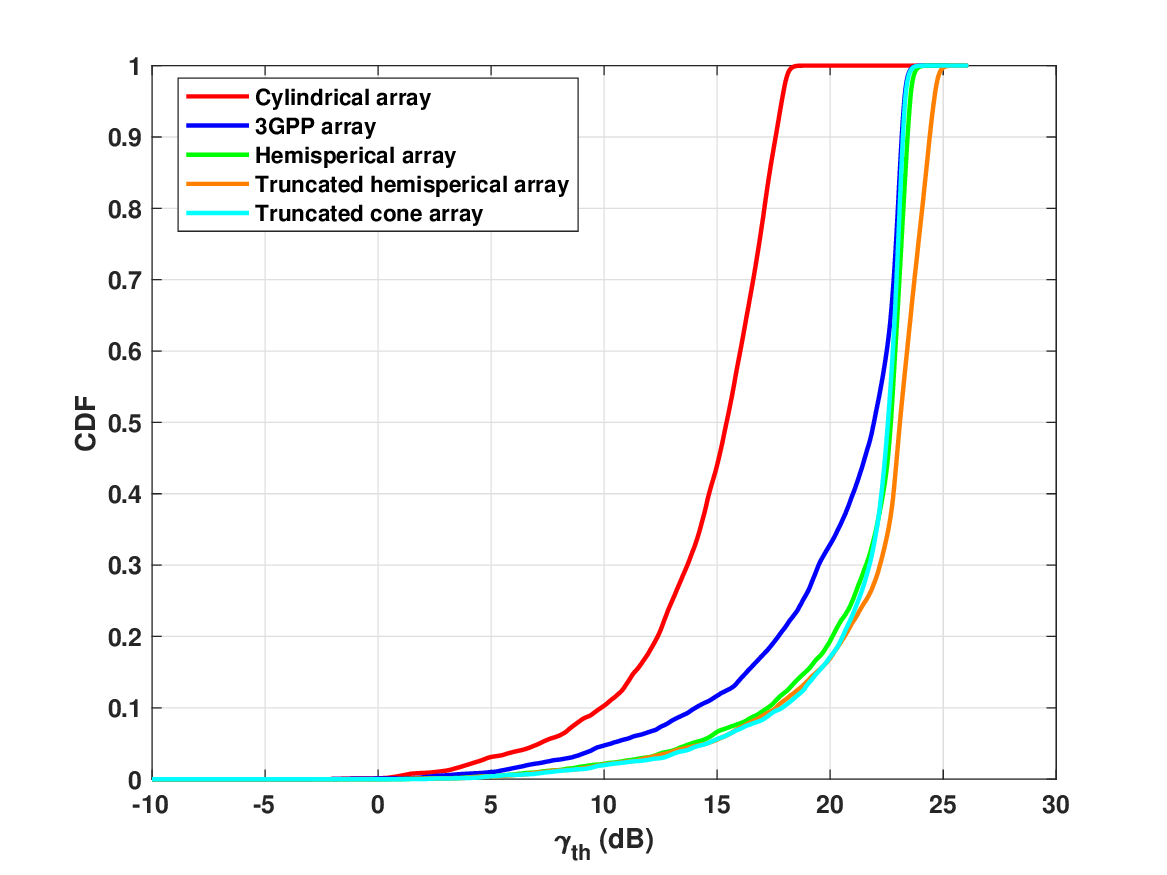}
    \label{Fig: RR20K20}
}
\subfigure[$K=50, R_{\text{radius}}=20$ km]{
    \includegraphics[width=2.2in]{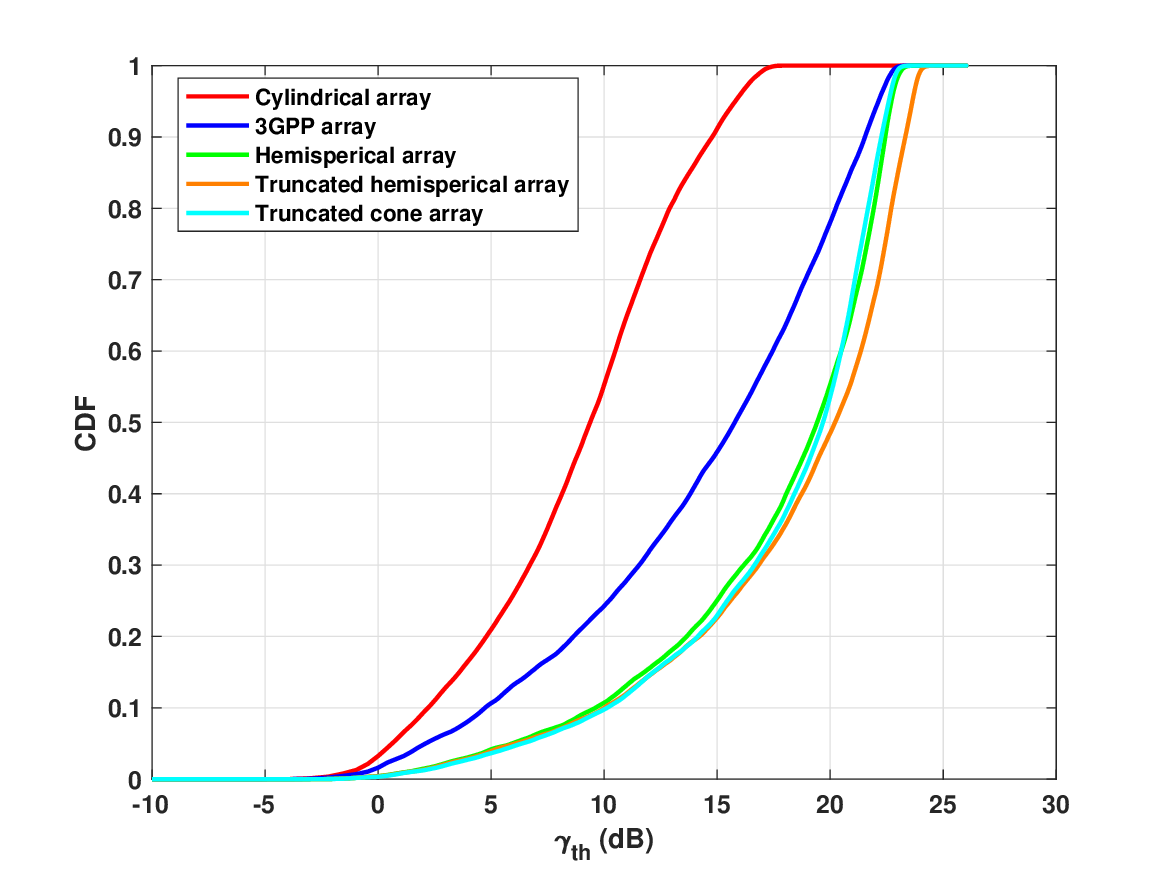}
    \label{Fig: RR20K50}
}
\caption{Uniformly distributed UEs in a circular cell of radius $R_{\text{radius}}=20$ km.}
\label{Fig:R20_uniform}
\end{figure*}
We assume that the UEs are uniformly distributed within a circular area of radius $R_{\text{radius}}$ km on the ground as illustrated in Fig.~\ref{Fig:R100_uniform} and Fig.~\ref{Fig:R20_uniform}. 
Specifically, Fig.~\ref{Fig:R100_uniform} and Fig.~\ref{Fig:R20_uniform} show the SINR distributions for different antenna arrays and different numbers of UEs within coverage areas of $R_{\text{radius}}=100$ km and $R_{\text{radius}}=20$ km, respectively.
It is observed that cylindrical array consistently exhibits inferior performance compared to other array types across all simulation scenarios. This performance degradation is mainly due to the fact that the boresights of antenna elements on the curved surface of the cylindrical array are parallel to the ground, resulting in a significant loss of antenna gain for ground UEs.
In addition, for a small number of UEs, the performance of the other four array types is generally comparable. However, as the number of UEs increases, the 3GPP array exhibits poorer performance than the hemispherical, truncated-cone, and truncated-hemispherical arrays.
This performance degradation can be attributed to the fact that, under high UE density, a higher degree of angular freedom in the azimuth domain is essential for effective interference mitigation and for increasing the probability that antenna elements are directly aligned with UEs. In contrast, the 3GPP array consists of only six side panels, which limits its azimuthal angular resolution and beamforming flexibility.
Further, when $R_{\text{radius}}=100$ km, it is noted that the hemispherical, truncated cone, and truncated hemispherical array exhibit similar performance.
However, when $R_{\text{radius}}=20$ km, 
the truncated hemispherical array achieves the best performance in terms of high-SINR regime. This is because, relative to the hemispherical configuration, the addition of panels at the bottom enables better coverage of UEs located directly beneath the HAPS. Moreover, compared with the truncated cone, the truncated hemispherical configuration has a higher probability of aligning with UEs due to its greater flexibility in both azimuth and zenith angles.
\par
\begin{figure*}[!h]
\centering
\subfigure[Standard deviation $\sigma=10$ km]{
    \includegraphics[width=2.2in]{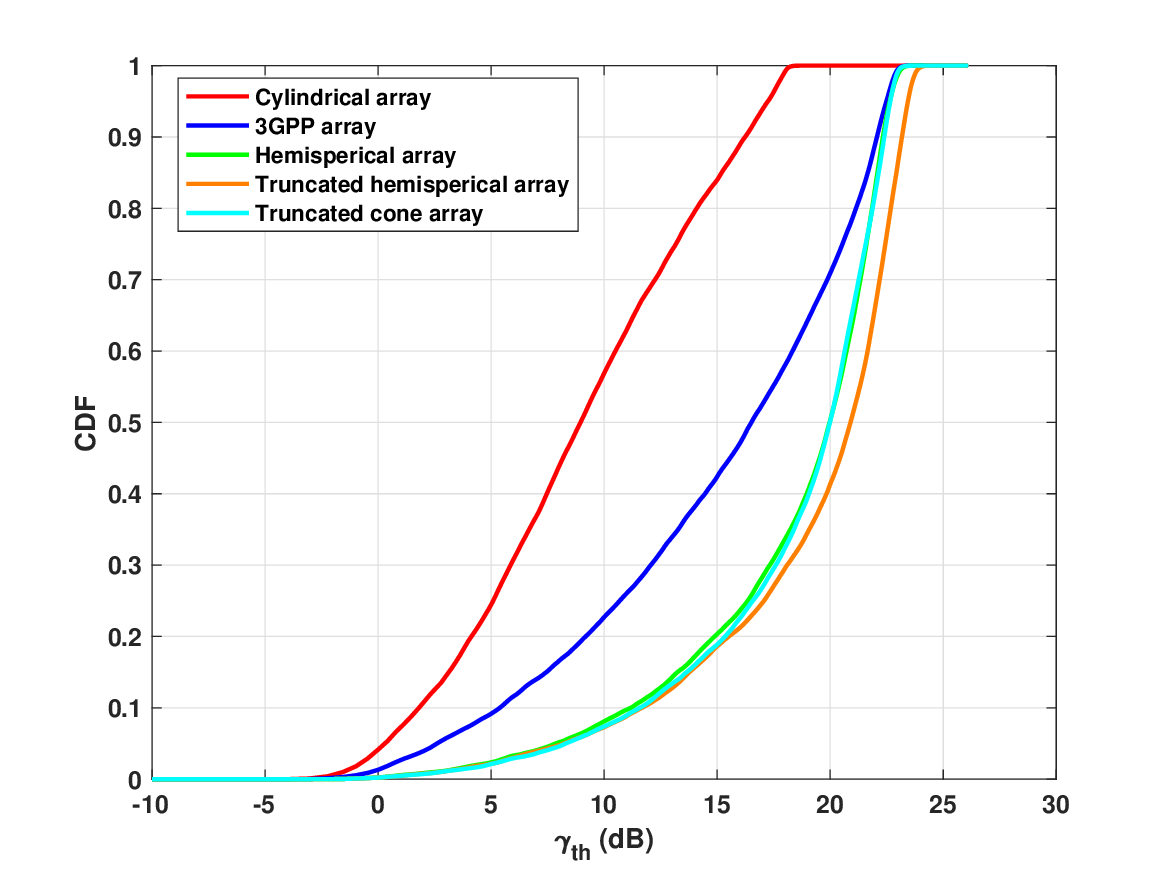}
    \label{Fig: sigma10}
}
\subfigure[Standard deviation $\sigma=20$ km]{
    \includegraphics[width=2.2in]{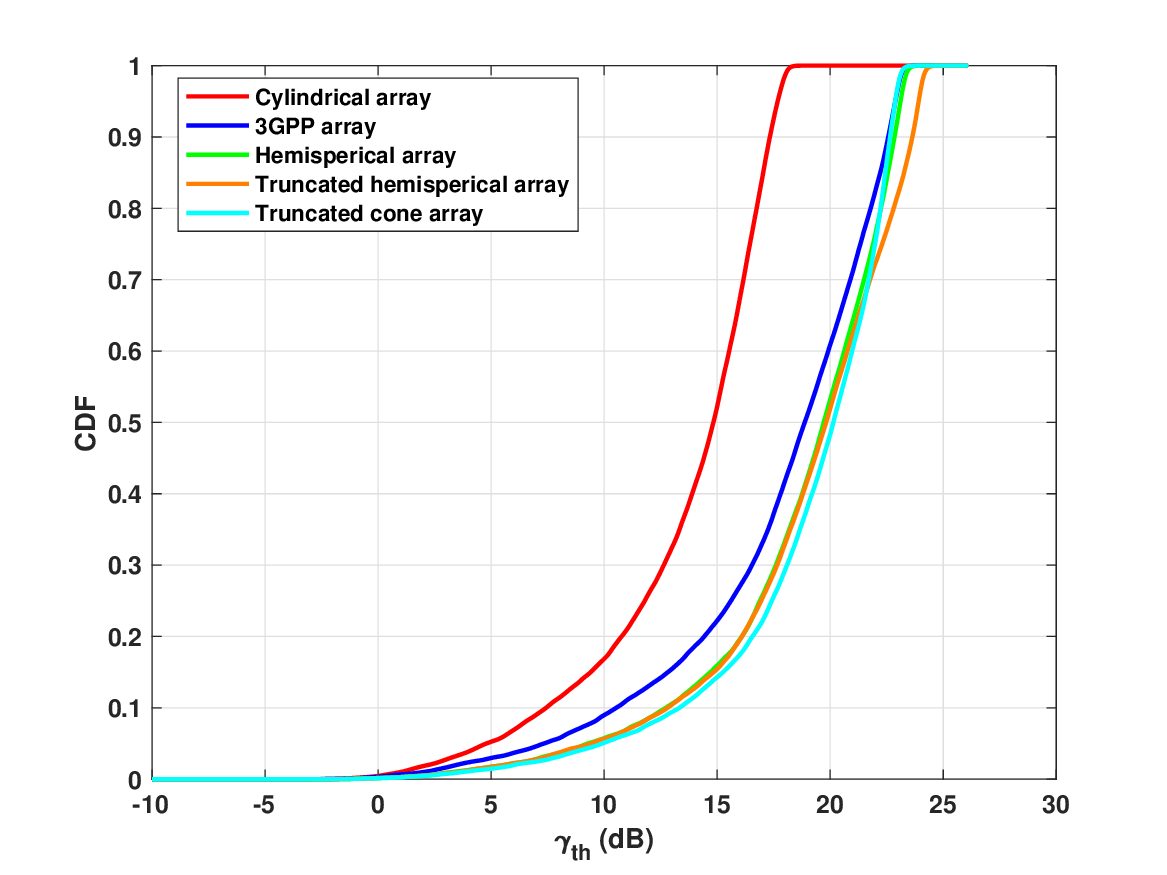}
    \label{Fig: sigma20}
}
\subfigure[Standard deviation $\sigma=50$ km]{
    \includegraphics[width=2.2in]{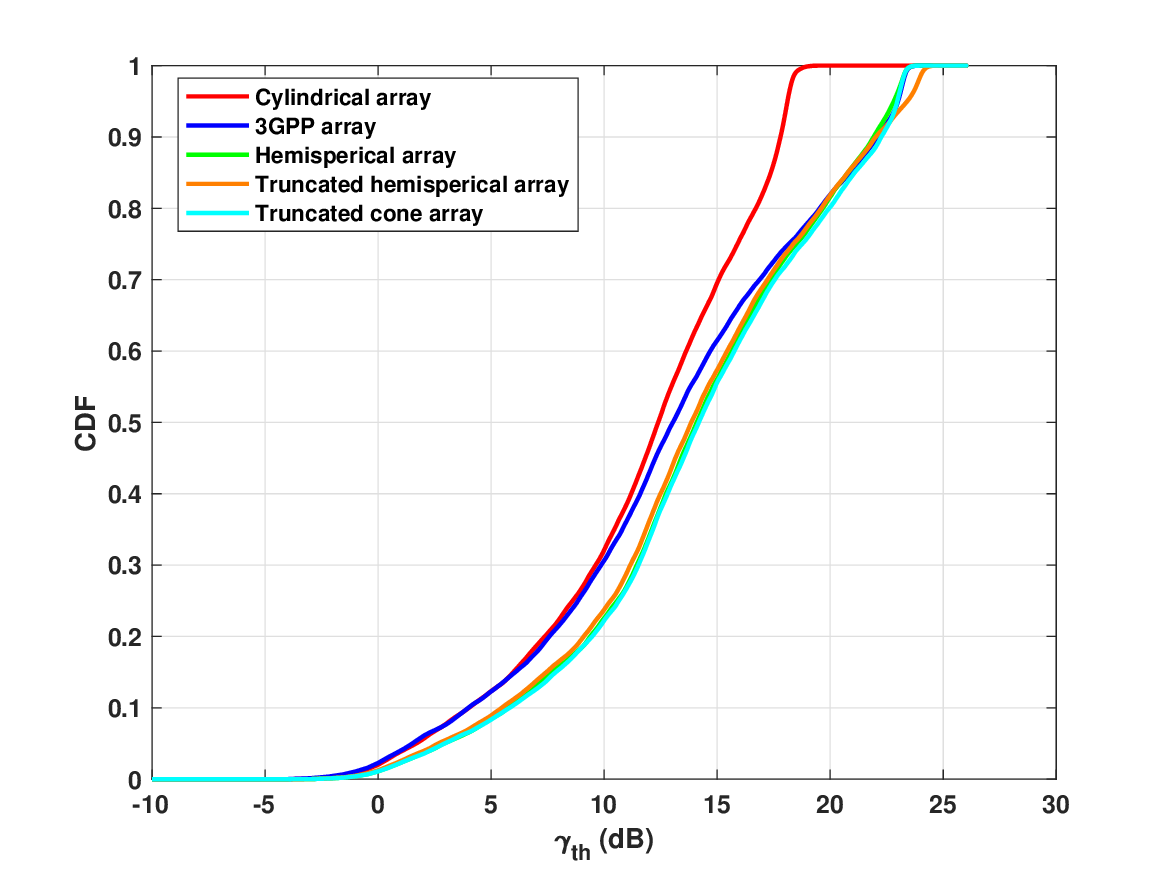}
    \label{Fig: sigma50}
}
\caption{Gaussian spatial distribution of UEs within a circular cell of radius $R_{\text{radius}}=100$ km.}
\label{Fig: R100_gaussian}
\end{figure*}

\begin{figure*}[!h]
\centering
\subfigure[Standard deviation $\sigma=10$ km]{
    \includegraphics[width=2.2in]{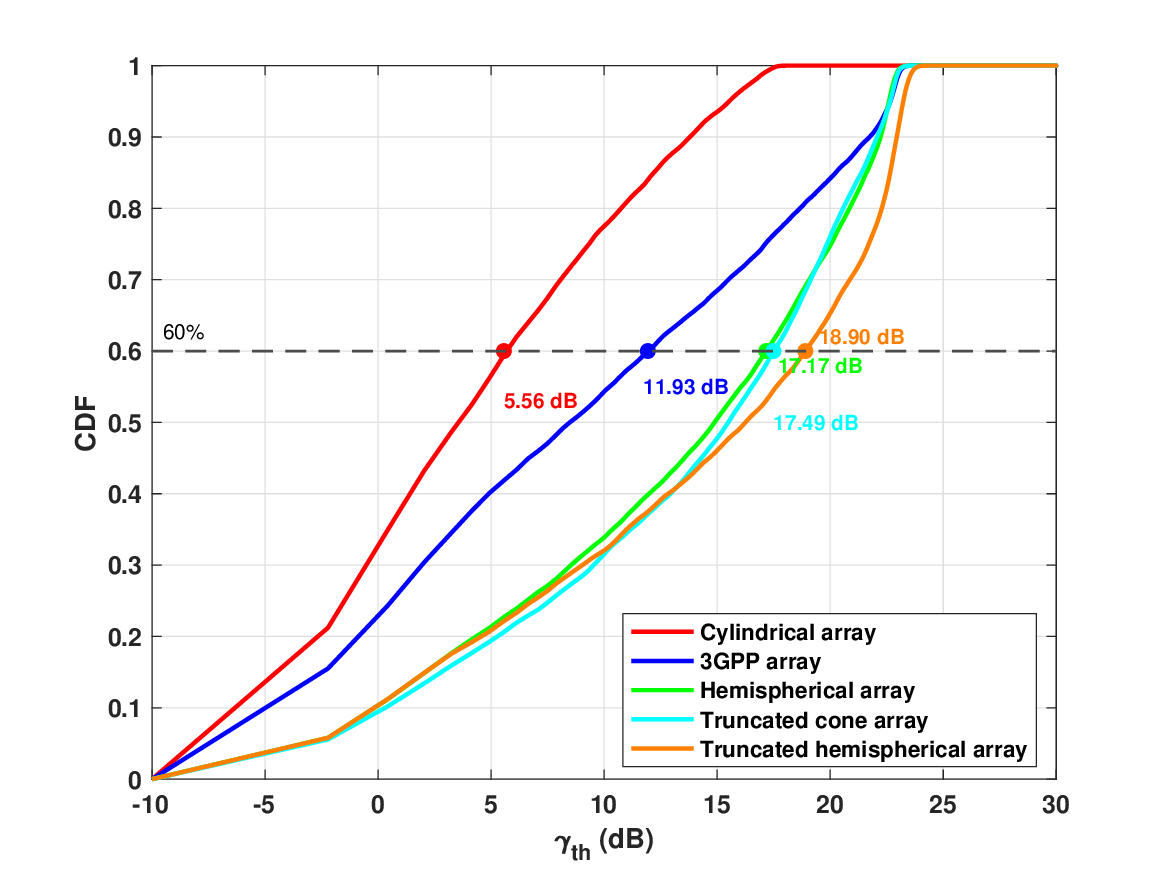}
    \label{Fig: sigma10_RR20}
}
\subfigure[Standard deviation $\sigma=20$ km]{
    \includegraphics[width=2.2in]{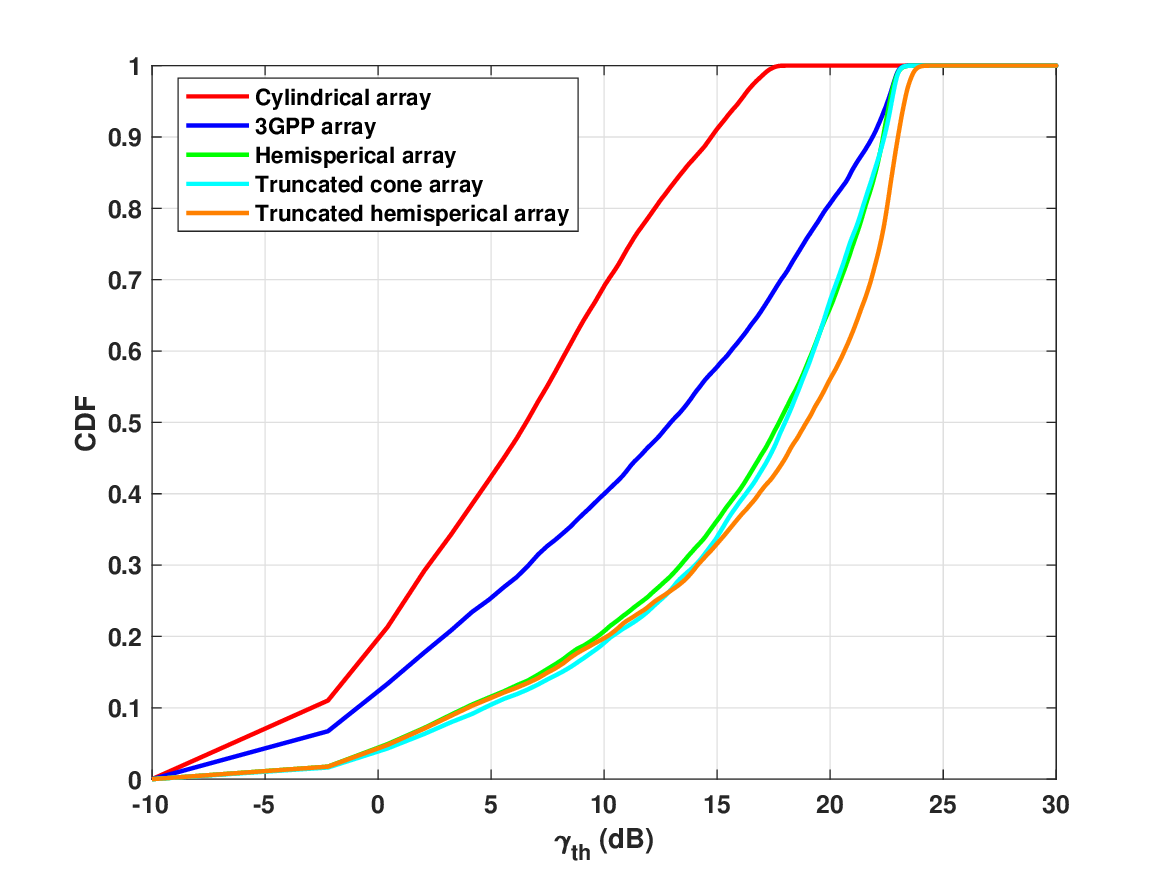}
    \label{Fig: sigma20_RR20}
}
\subfigure[Standard deviation $\sigma=50$ km]{
    \includegraphics[width=2.2in]{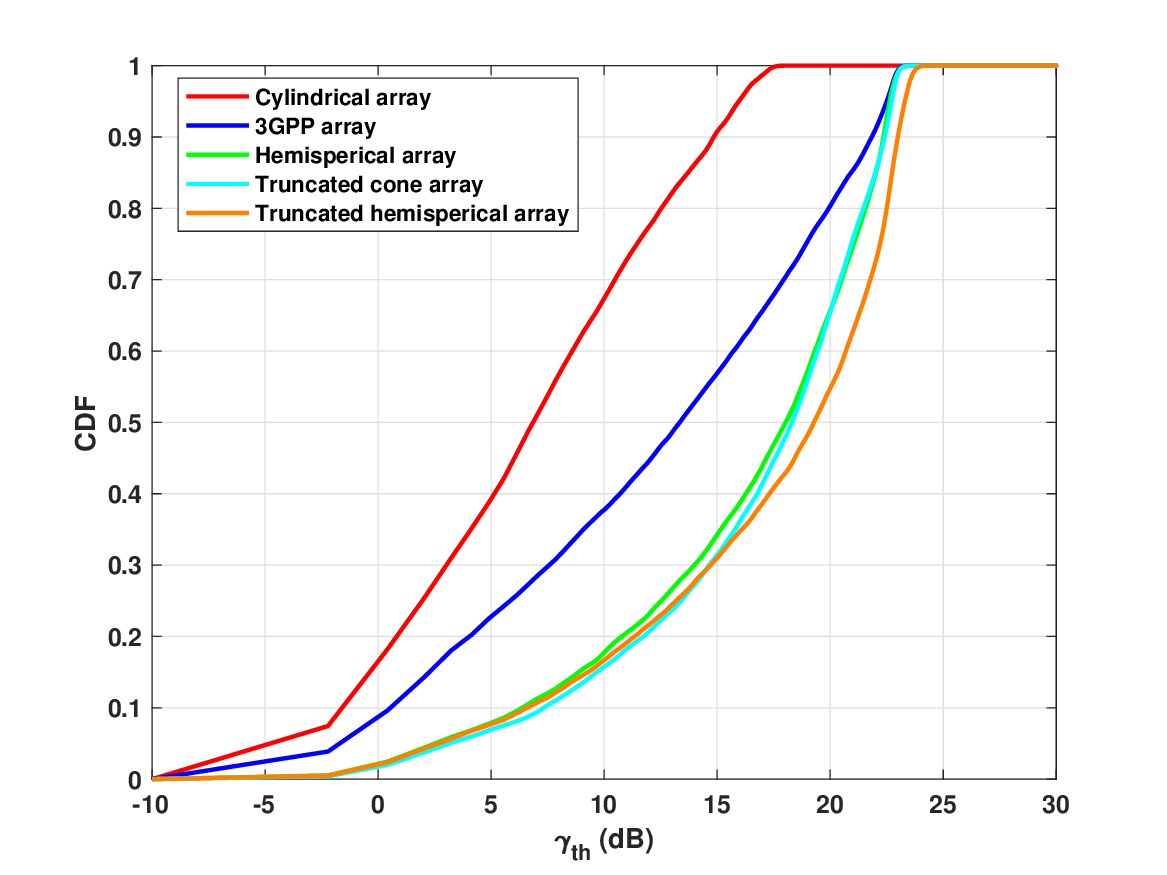}
    \label{Fig: sigma50_RR20}
}
\caption{Gaussian spatial distribution of UEs within a circular cell of radius $R_{\text{radius}}=20$ km.}
\label{Fig: R20_gaussian}
\end{figure*}

We then investigate in which the UEs are Gaussian distributed within a circular area of radius $R_{\text{radius}}$ km on the ground, with the number of UEs $K=50$.
Specifically, the UE location $\mathbf{X}=[X,Y]^{\mathsf T}\in\mathbb{R}^2$ is modeled as a two-dimensional isotropic 
Gaussian random vector, i.e., $\mathbf{X}\sim\mathcal{N}(\mathbf{0},\sigma^2\mathbf{I}_2)$, and truncated to the circular 
region $\mathcal{D}=\{\mathbf{x}:\|\mathbf{x}\|\le R_{\text{radius}}\}$. Let $(\tilde r,\tilde\theta)$ denote the polar 
coordinates of $\mathbf{X}$, where $\tilde r=\|\mathbf{X}\|$ and $\tilde\theta=\mathrm{atan2}(Y,X)$. Conditioning on the 
truncation event $\|\mathbf{X}\|\le R_{\text{radius}}$, the joint PDF of $(\tilde r,\tilde\theta)$ is given by
$$
f_{\tilde r,\tilde\theta \mid \|\mathbf{X}\|\le R_{\text{radius}}}(\tilde r,\tilde\theta)
=\frac{\frac{1}{2\pi\sigma^2}\, \tilde r\, \exp\!\left(-\frac{\tilde r^2}{2\sigma^2}\right)}
{F(R_{\text{radius}})},
$$
where the normalization constant is
\[
F(R_{\text{radius}})=1-\exp\!\left(-\frac{R_{\text{radius}}^2}{2\sigma^2}\right),
\]
with $0\le \tilde r\le R_{\text{radius}}$ and $\tilde\theta\in[0,2\pi)$. 
\par
Fig.~\ref{Fig: R100_gaussian} and Fig.~\ref{Fig: R20_gaussian} illustrate the SINR performance under Gaussian user distributions with coverage radii of $100$ km and $20$ km, respectively. It can be observed that when the user distribution is highly concentrated (i.e., $\sigma = 10$ km), the proposed truncated hemispherical array achieves the best performance in the high-SINR regime. This is because most UEs are located near the center of the coverage area, allowing the truncated hemispherical array to fully exploit its enhanced angular flexibility and the additional bottom panel to provide higher antenna gain toward the HAPS--UE links.
For the $100$ km coverage radius, as the standard deviation increases, the performance gap among the truncated hemispherical, hemispherical, truncated cone, and hexagonal arrays gradually diminishes, and these architectures achieve comparable SINR performance. This is because a larger standard deviation spreads UEs over a wider area, reducing the advantage of the additional downward-facing elements. In contrast, for the $20$ km coverage radius, the truncated hemispherical array consistently achieves the best performance across different standard deviations. Since the entire coverage area remains close to the HAPS nadir, the additional bottom panel continues to provide a significant gain advantage, enabling the proposed architecture to maintain superior SINR performance even as the user distribution becomes more dispersed.
\par
\begin{figure*}[!h]
\centering
\subfigure[$N_p=1$]{
    \includegraphics[width=2.2in]{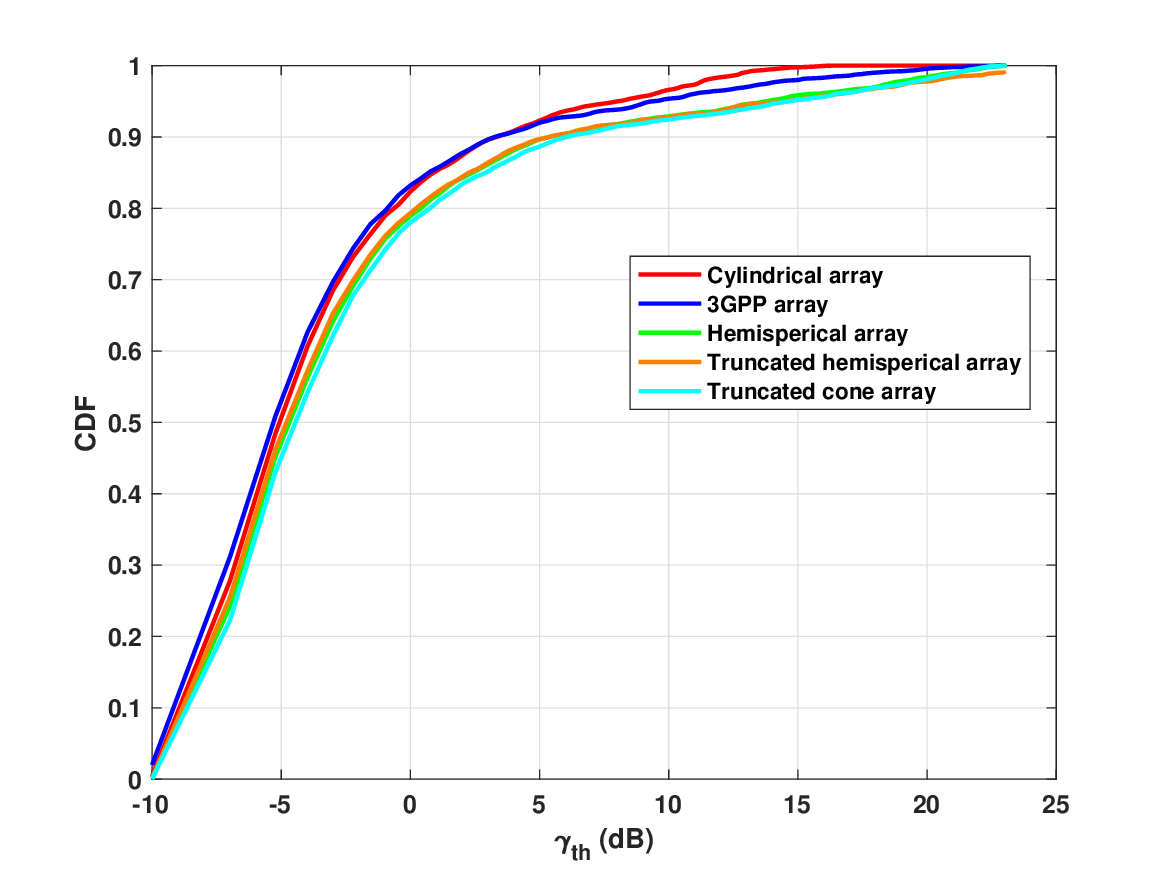}
    \label{Fig: PCP_N1}
}
\subfigure[$N_p$=2]{
    \includegraphics[width=2.2in]{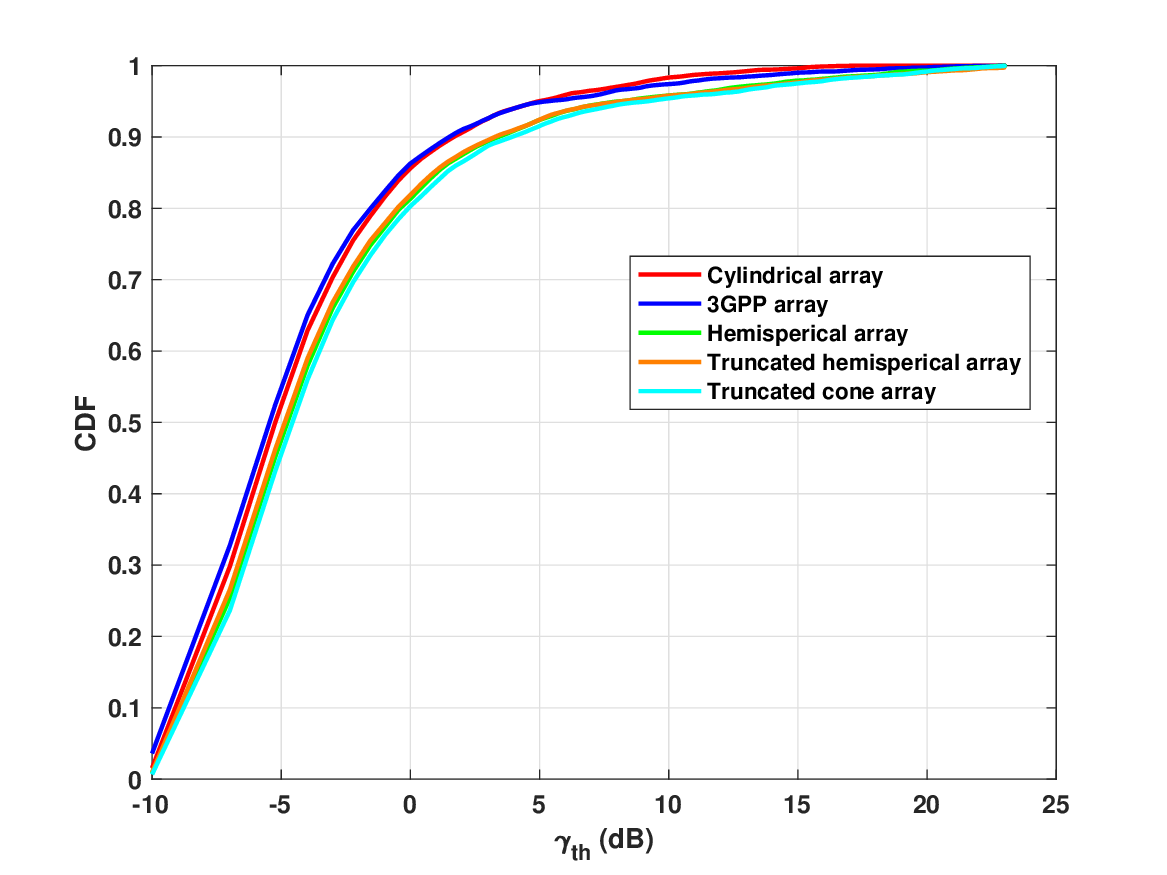}
    \label{Fig: PCP_N2}
}
\subfigure[$N_p=3$]{
    \includegraphics[width=2.2in]{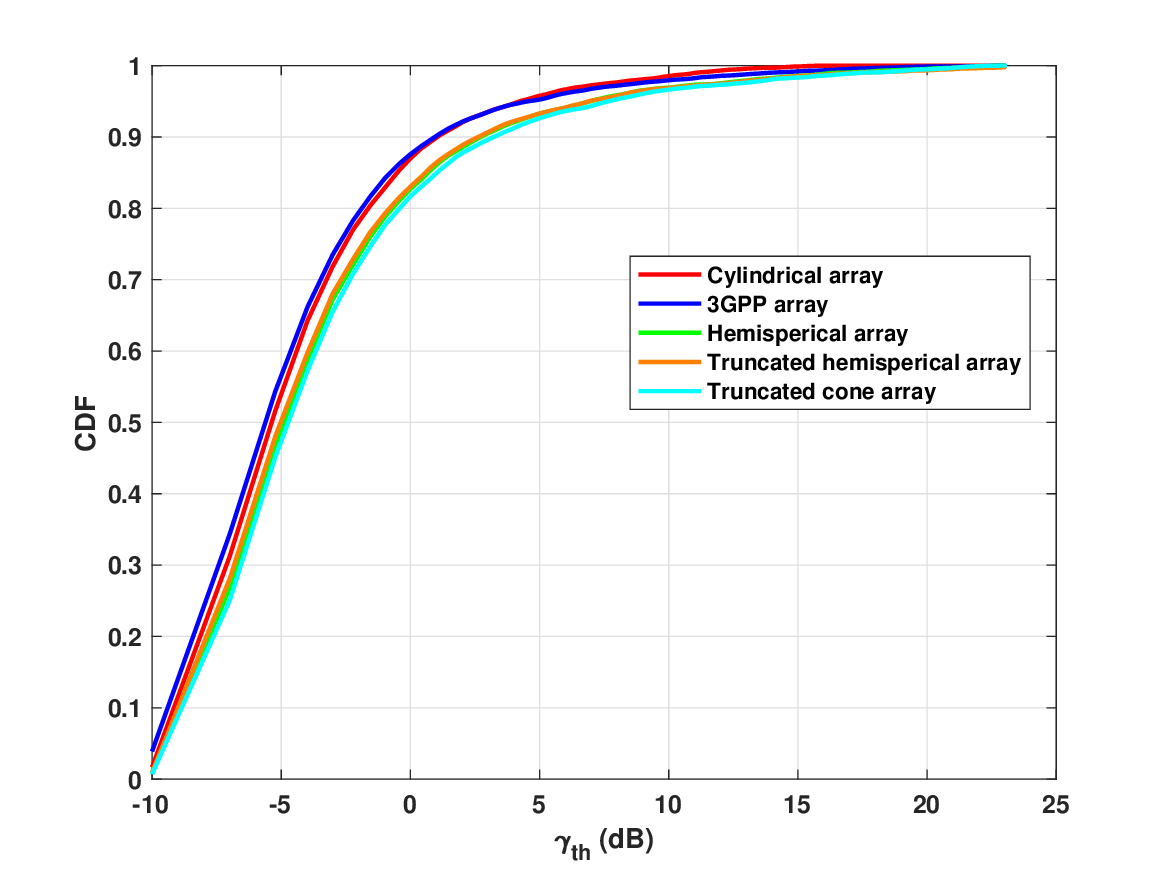}
    \label{Fig: PCP_N3}
}
\caption{PCP distribution of UEs within a circular cell of radius $R_{\text{radius}}=100$ km.}
\label{Fig: R100_PCP}
\end{figure*}
Finally, to model spatially clustered UE distributions resembling an archipelago, a Poisson Cluster Process (PCP) is adopted. In this model, $N_p$
 parent points, representing islands, are first generated according to a uniform distribution. $20$ UEs are then uniformly distributed around each parent point, forming local clusters with a radius of $R_{\rm cluster}=10$ km, as illustrated in Fig.~\ref{Fig: R100_PCP}, for different values of $N_p$. 
In all simulations, it is observed that the hemispherical, truncated cone, and truncated hemispherical arrays exhibit similar performance and outperform the cylindrical and 3GPP arrays. Among them, the truncated cone array performs slightly better than the others.
\subsection{Discussion and Recommendation}
Based on the above results, the proposed truncated cone and truncated hemispherical arrays achieve performance comparable to that of the hemispherical array. Moreover, as shown in Fig.~\ref{Fig:R20_uniform} and Fig.~\ref{Fig: sigma10}, the proposed truncated hemispherical array is well suited to scenarios with either a small coverage area or a large coverage area where UEs are densely distributed.
In addition, compared with the hemispherical array, the truncated cone array has a more regular geometric structure, which enables antenna elements with similar orientations to be efficiently grouped into subarrays. This structural advantage allows multiple antenna elements to share a single RF chain in a hybrid beamforming architecture, thereby reducing the required number of RF chains. Consequently, the truncated cone array is recommended for other deployment scenarios

\section{Conclusion}
This paper investigated the uplink performance of HAPS systems with different
antenna array configurations in different UE distributions.
Simulation results demonstrate that antenna geometry and orientation have a
significant impact on user SINR under various user distributions.
The proposed array designs achieve performance comparable to or better than
existing solutions across multiple deployment scenarios, thereby providing
useful insights for the design of future high-capacity HAPS-enabled networks.

\appendices
\section{}
\begin{lemma}[Coordinates System and Array Orientation]
\label{lemma: GCS2LCS}
Let a Global Coordinate System (GCS) be defined by coordinates $(x,y,z,\theta,\phi)$ and a Local Coordinate System (LCS) defined by coordinates $(x_a,y_a, z_a, \theta_a, \phi_a)$.
Considering an arbitrary 3D-rotation of the LCS with respect to the
GCS given by the angles $\alpha, \beta, \gamma$, where $\alpha$ is the bearing angle, $\beta$ is the downtilt angle, and $\gamma$ is the slant angle. 
The rotation matrix is given by \cite{3gpp_tr38901}
\begin{equation}
\begin{aligned}
\mathbf{R}=&\mathbf{R}_Z(\alpha)\mathbf{R}_Y(\beta)\mathbf{R}_X(\gamma)\\
=&\begin{bmatrix}
\cos\alpha & -\sin\alpha & 0 \\
\sin\alpha & \cos\alpha  & 0 \\
0          & 0           & 1
\end{bmatrix}
\begin{bmatrix}
\cos\beta & 0 & \sin\beta \\
0         & 1 & 0 \\
-\sin\beta& 0 & \cos\beta
\end{bmatrix}\\
\times&
\begin{bmatrix}
1 & 0          & 0 \\
0 & \cos\gamma & -\sin\gamma \\
0 & \sin\gamma & \cos\gamma
\end{bmatrix}
\end{aligned}
\end{equation}
Define the unit vector in GCS by the spherical coordinates with $(\theta,\phi)$. The Cartesian representation of that point is given by
$$
\mathbf{e}_r=(\sin\theta\cos\phi, \sin\theta\sin\phi, \cos\theta)
$$.
Hence, the local angles $\theta_a$ and $\phi_a$ is given by 
$$
\theta_a=\arccos\!\left(
\begin{bmatrix}
0 \\ 0 \\ 1
\end{bmatrix}^{\!T}
\mathbf{R}^{-1}\mathbf{e}_r
\right),
$$
and 
$$
\phi_a
= \arg\!\left(
\begin{bmatrix}
1 \\ j \\ 0
\end{bmatrix}^{\!T}
R^{-1}\mathbf{e}_r
\right),
$$
where $\mathbf{R}^{-1}=\mathbf{R}^T$.
\end{lemma}

\section{Antenna Element Radiation Patterns}
We adopt the antenna radiation patterns as defined in the 3GPP specifications \cite{3gpp_ntn_rf_r18}, 
where the overall antenna gain is obtained as the combination of the zenith and azimuth patterns:
\begin{equation}
    A_E(\phi, \theta)=G_{E,\text{max}}-\min\{-[A_{E,H}(\phi)+A_{E,V}(\theta)],A_{\text{m}}\} ,
\end{equation}
where,
$
A_{E, V}\left(\theta\right)=-\min \left\{12\left(\frac{\theta-90^{\circ}}{\theta_{3 \mathrm{dB}}}\right)^2, S L A_V\right\} \mathrm{dB},
$
and 
$
A_{E, H}\left(\phi\right)=-\min \left\{12\left(\frac{\phi}{\phi_{3 \mathrm{dB}}}\right)^2, A_m\right\} \mathrm{dB}.
$
Here, the azimuth angle $\phi$ and the zenith angle $\theta$ are defined with respect to the 
local spherical coordinate system of the antenna.
whereas $A_{E,H}(\phi)$ and $A_{E,V}(\theta)$ represent the horizontal and vertical element radiation patterns, 
$G_{E,\text{max}}$ is the maximum directional gain of the antenna element, 
$\phi_{3\mathrm{dB}}$ and $\theta_{3\mathrm{dB}}$ are the half-power beamwidths in the azimuth and elevation planes, respectively, 
$SLA_{V}$ denotes the vertical side-lobe attenuation, and $A_m$ is the maximum attenuation.
The global field pattern of a vertically polarized antenna element in linear scale is $\sqrt{A_E(\phi, \theta)|_{\text{lin}}}=\sqrt{10^{\frac{G_{E,\text{max}}}{10}} g_E(\phi,\theta)}$, where $g_E(\phi,\theta) \approx g_{E,V}(\theta)g_{E,H}(\phi)$ with
\begin{equation}
\label{Eq:G_EV}
     g_{E,V}(\theta)=\exp{\left(-1.2\left(\frac{\theta-90^{\circ}}{\theta_{3 \mathrm{dB}}}\right)^2 \ln{10}\right)},
\end{equation}
\begin{equation}
\label{Eq:G_EH}
    g_{E,H}(\phi)=\exp{\left(-1.2\left(\frac{\phi}{\phi_{3 \mathrm{dB}}}\right)^2 \ln{10}\right)}.
\end{equation}
\par

\bibliography{my_bibliography}
\bibliographystyle{IEEEtran}
\end{document}